\RequirePackage{fix-cm}

\documentclass[twocolumn,epjc3]{svjour3}
\usepackage{amsmath}
\usepackage[english]{babel}
\usepackage{graphicx}
\usepackage{psfrag}
\usepackage{cite}

\newcommand{\ud}{\mathrm{d}}

\smartqed  
\RequirePackage{graphicx}
\journalname{Eur. Phys. J. C}
\begin{document}
\title{BLM Scale Fixing in Event Shape Distributions}
\author{Thomas~Gehrmann\thanksref{e1,addr1},
  Niklaus~H\"afliger\thanksref{e2,addr1} \and
  Pier~Francesco~Monni\thanksref{e3,addr1,addr2,addr3} 
}

\thankstext{e1}{e-mail: thomas.gehrmann@uzh.ch}
\thankstext{e2}{e-mail: niklaus@lcc.ch}
\thankstext{e3}{e-mail: pfmonni@physik.uzh.ch}

\institute{Physik-Institut, Universit\"at Z\"urich,
Winterthurerstrasse 190, CH-8057 Z\"urich, Switzerland\label{addr1}
\and
Rudolf Peierls Centre for Theoretical Physics, University
  of Oxford, 1 Keble Road, Oxford OX1 3NP, UK\label{addr2}
\and
Institute for Particle Physics Phenomenology, University of Durham,
 Durham DH1 3LE, UK\label{addr3}}

\date{Received: date / Accepted: date}

\maketitle

\begin{abstract}
  We study the application of the 
  Brodsky-Lepage-Mackenzie (BLM) scale setting prescription to event shape
  distributions in electron-positron collisions. The renormalization
  scale is set dynamically according to the BLM method. We study NLO
  predictions and we discuss extensions of the prescription to NNLO.

\end{abstract}

\section{Introduction}
\label{intro}

A key problem in making precise predictions in perturbative QCD
concerns the choice of the renormalization scale for a process and the
theoretical uncertainty associated with it. The dependence of the
perturbative result upon such a scale gives an estimate of the size of
unknown higher-order perturbative corrections. Since in general only
the first few terms of the perturbative series are actually known, it
is of primary relevance to figure out how to choose the
renormalization scale in order to minimize missing terms. Different
possible prescriptions have been proposed in the literature, {\it
  e.g.} fastest apparent convergence~\cite{Grunberg:1980ja}, principle
of minimum sensitivity (PMS)~\cite{Stevenson:1980du} and the
Brodsky-Lepage-Mackenzie (BLM) method~\cite{Brodsky:1982gc}. Commonly
the scale is set to some characteristic scale of the process, which
can depend on the observables under consideration. In the present work
we apply the BLM method to event-shape distributions in
electron-positron collisions. We analyze the standard set of six
event-shape observables described in e.g.\ \cite{Jones:2003yv}, which
have been measured precisely at $e^+e^-$
colliders~\cite{aleph,opal,delphi,tasso,l3,jade,sld}.  Perturbative
QCD predictions up to
NNLO~\cite{GehrmannDeRidder:2007hr,Weinzierl:2009ms} and electro-weak
corrections up to NLO~\cite{Denner:2009gx} are available for these
observables.  Moreover, resummations of Sudakov logarithms have been
derived for these observables at different logarithmic
accuracies~\cite{CTTW,Catani:1998sf,broadenings,Becher:2011pf,y3,Becher:2008cf,Monni:2011gb,Chien:2010kc,Becher:2012qc}.

The BLM method was initially formulated for next-to-leading order
(NLO) predictions. For differential quantities, the BLM prescription
results in a dynamical renormalization scale set on a bin-by-bin
basis. We compare different choices of the renormalization scale and
we analyze the extension of the method to NNLO. The resulting
predictions are compared to experimental data from the ALEPH
collaboration~\cite{aleph}. The paper is organized as follows. In
Section~\ref{sec:framework} we report the theoretical framework. In
Section~\ref{sec:blmnlo} we recall the BLM method and discuss its
implementation for event-shape variables and the extension of the
approach to NNLO. Numerical results are reported in
Section~\ref{sec:results}, while Section~\ref{sec:Conclusions}
contains our conclusions.

\section{Event shapes in perturbative QCD}
\label{sec:framework}
For the set of event-shape observables analyzed here, NNLO predictions
were computed in
refs.~\cite{GehrmannDeRidder:2007hr,Weinzierl:2009ms}. We can express
their perturbative expansion in the form
\begin{align}
  {1\over \sigma} {\ud \sigma \over \ud y} &=A(y) \left (
    {\alpha_s(\mu_1) \over 2\pi} \right ) +   B(y,\mu_1)  \left
    ({\alpha_s(\mu_2) \over 2\pi} \right )^2\notag \\
  \label{eq:eventshape}
  &+  C(y,\mu_1,\mu_2)  \left ( {\alpha_s(\mu_3) \over 2\pi} \right )^3 + ...,
\end{align}
where $y$ stands for any event-shape variable and the dots indicate
missing higher-order corrections of $\mathcal{O}(\alpha_s^4)$, with
$\sigma$ being the total cross section. The renormalization scales
$\mu_1$, $\mu_2$ and $\mu_3$ indicate the scales at which the coupling
constant is evaluated in the leading, next-to-leading and
next-to-next-to-leading order corrections, respectively. The strong
coupling in Eq.~\eqref{eq:eventshape} is commonly evaluated at some
renormalization scale $\mu$ of the order of the
centre-of-mass energy $Q$, {\it i.e.}
$\mu_1=\mu_2=\mu_3=Q$.

For our analysis it is useful to express explicitly the dependence of
the perturbative coefficients in Eq.~\eqref{eq:eventshape} on the
number of active quark flavours $n_F$. We thus write
\begin{align}
A(y) & =  A_0(y)\,,\notag\\
B(y,Q) & =  B_0(y) + B_1(y) n_F\,,\notag\\
\label{eq:pertcoeffs}
C(y,Q,Q) & =  C_0(y) + C_1(y) n_F + C_2(y) n_F^2\,,
\end{align}
where the coefficients $A_0$, $B_0$, $B_1$, $C_0$, $C_1$ and $C_2$ are
evaluated at a fixed renormalization scale $Q$.

\section{\bf{The BLM method}}
\label{sec:blmnlo}
In this section we briefly recall the BLM scale setting method
introduced in~\cite{Brodsky:1982gc}. The method gives a simple
prescription to set the renormalization scale for a process, in order
to improve the convergence of the perturbative expansion.  The main
idea is to redefine the coupling constant such that all contributions
arising from corrections to gauge boson propagators are absorbed into
it.  In QED, the running of the coupling is exclusively ruled by vacuum
polarization insertions in the photon propagator. The latter diagrams
are automatically absorbed into the QED coupling constant
$\alpha(k^2)$ through the photon wave function renormalization
\begin{align}
\label{eq:wavefunction}
\frac{1}{-k^2-i0}\to \frac{\alpha(-k^2)}{-k^2-i0},
\end{align}
which fully defines the QED running
coupling. Eq.~\eqref{eq:wavefunction} shows that the absorption of
vacuum polarization diagrams into the coupling makes the latter run
with the virtuality of the virtual photon. This provides us with a
prescription to choose the coupling scale when evaluating Feynman
diagrams. A direct consequence of this prescription is that each
Feynman diagram contributing to a given amplitude will have a
different scale at which the coupling must be evaluated. At higher
orders, where loop integrals are present, this prescription would make
the loop integration quite cumbersome and thus it is not practical. It
is then customary to choose a common renormalization scale at which
all the couplings $\alpha$ present in the process are evaluated.  If
one considers using the mean value theorem to evaluate the resulting
loop integral, there must be some momentum scale $k^2 = Q^{*2}$ which
dominates the integral, thus minimizing higher-order
corrections. Higher-order corrections would naturally require a
different scale, leading to the perturbative expansion
\begin{align}
{1\over \sigma} {\ud \sigma \over \ud y} &=A(y) \left ( {\alpha(Q^*) \over 2\pi} \right ) +   B(y,Q^*)  \left ( {\alpha(Q^{**}) \over 2\pi} \right )^2\notag\\
&+  C(y,Q^*,Q^{**})  \left ( {\alpha(Q^{***}) \over 2\pi} \right )^3 + ...,
\label{eq:eventshapeBLM}
\end{align}
for a generic observable $y$.

The BLM method suggests to fix the scales $Q^*$, $Q^{**}$, $Q^{***}$,
$\ldots$ so that to absorb the vacuum polarization contributions into
the coupling at each order in perturbation theory. In QED, at low
orders we can easily identify those contributions with the
$n_F$-dependent terms in the perturbative expansion, with $n_F$ being
the number of active flavours.  At higher orders, additional
$n_F$-dependent terms arise through fermion boxes which are UV finite
and thus must not be absorbed into the coupling.

The extension of this prescription to QCD is not trivial, since more
diagrams (gluon and ghost loops) contribute to the running of the
strong coupling constant. Ref.~\cite{Brodsky:1982gc} suggests to
implement the same prescription used in QED, but with the replacement
\begin{equation}
\label{eq:QEDtoQCD}
\beta^{\rm QED}\to \beta
\end{equation}
where $\beta$ denotes the QCD $\beta$ function.  Once again this
amounts to absorb the vacuum polarization diagrams into the strong
coupling constant, and to set the renormalization scales such that
$n_F$-dependent terms vanish at each order in perturbation
theory. This recipe will work unless fermionic box diagrams are
present. In this case, it is not possible to disentangle the vacuum
polarization contributions from the remaining $n_F$ dependent terms
and the prescription does not apply. In the process $e^+e^-\to$\,jets
that we want to study, such terms only appear at and beyond NNLO, so
the BLM method can be applied at NLO. The prescription outlined
in~\eqref{eq:QEDtoQCD} implies that also the $\sim C_A$ contributions
to vacuum polarization diagrams (due to gluon and ghost loops) are
absorbed into the running of $\alpha_s$.  The final perturbative
expansion will be free of vacuum polarization diagrams which are
responsible for the leading renormalon growth, i.e. $\sim
\alpha_s^{n+1} \beta_0^n n!$ (see e.g. ref.~\cite{Beneke:1998ui}), so
it is expected to have better convergence properties.

The LO BLM renormalization scale $Q^*$ can be obtained by cancelling
the $n_F$ dependence of the NLO coefficient in the expansion
\begin{align}
{1\over \sigma} {\ud \sigma \over \ud y} &=A(y) \left ( {\alpha_s(Q^*) \over 2\pi} \right ) +   B(y,Q^*)  \left ( {\alpha_s(Q^{**}) \over 2\pi} \right )^2\notag\\
&+  C(y,Q^*,Q^{**})  \left ( {\alpha_s(Q^{***}) \over 2\pi} \right )^3 + ...,
\end{align}
 where the perturbative coefficients read
(we set $C_A=3$, $C_F=4/3$ and $T_F=1/2$)
\begin{align}
 A(y)&  =  A_0(y)\,,\notag\\
 B(y,&Q^*) = B_0(y) + {11 \over 2} A_0(y) \log{Q^{*2} \over Q^2} \notag\\
 &+ \left ( B_1(y) - {1 \over 3} A_0(y) \log {Q^{*2} \over Q^2} \right) n_F\,,  \notag\\
 C(y,&Q^*,Q^{**})  =  C_0(y)
    + {51 \over 2} A_0(y)\ln{Q^{*2} \over Q^2} \notag\\
 & - {121\over 4}A_0(y)\ln^2{Q^{*2} \over Q^2}
 +{121\over 2}A_0(y)\ln{Q^{*2} \over Q^2}\ln{Q^{**2} \over Q^2}\notag\\
 &+ 11 B_0(y)\ln{Q^{**2} \over Q^2}\notag\\
 &+ n_F\bigg[C_1(y)-{19\over 6}A_0(y)\ln{Q^{*2} \over Q^2}\notag\\
 &+{11\over 3}A_0(y)\ln^2{Q^{*2} \over Q^2} -{22\over 3}A_0(y)\ln{Q^{*2} \over Q^2}\ln{Q^{**2} \over Q^2} \notag\\
 &+\left(-{2\over 3}B_0(y)+11B_1(y)\right)\ln{Q^{**2} \over Q^2}\bigg]\notag\\
 \label{eq:expcoeffs}
 &+n_F^2\bigg[ C_2(y)-{1\over 9}A_0(y)\ln^2{Q^{*2} \over Q^2}\notag\\
 &+{2\over 9}A_0(y)\ln{Q^{*2} \over Q^2}\ln{Q^{**2} \over Q^2}-{2\over 3}B_1(y)\ln{Q^{**2} \over Q^2}\bigg].
\end{align}
Using the running coupling
expression we find
\begin{align}
\label{eq:BLMNLO}
Q^*= Q \exp\left\{\frac{3 B_1(y)}{2 A_0(y)}\right\},
\end{align}
where $Q$ is the centre-of-mass energy of the process. This amounts to
a dynamical scale which is set on a bin-by-bin basis. Considering the
expressions reported in Eq.~\eqref{eq:pertcoeffs}, the resulting
perturbative series reads
\begin{align}
\label{eq:BLMNLOseries}
{1\over \sigma} {\ud \sigma \over \ud y} &=A_0(y) \left ( {\alpha_s(Q^*) \over 2\pi} \right ) \notag\\
&+   \left(\frac{33}{2}B_1(y)+B_0(y)\right) \left ( {\alpha_s(Q^{**}) \over 2\pi} \right )^2 + ....
\end{align}
The NLO scale $Q^{**}$ is arbitrary at this order, and it is set by
higher-order vacuum polarization diagrams.

\subsection{\bf Extension to higher orders}
\label{sec:blmnnlo}
In the previous section we recalled the BLM scale fixing method which
led to the perturbative expansion Eq.~\eqref{eq:BLMNLOseries}, with
the renormalization scale set by Eq.~\eqref{eq:BLMNLO}.  To extend the
method to NNLO we have to set the LO and NLO scales ($Q^*$ and $Q^{**}$)
in order to absorb all vacuum polarization insertions up to ${\cal
  O}(\alpha_s^3)$. This is a non-trivial problem since beyond NLO
$n_F$-dependent terms arise also from UV-finite Feynman diagrams. At
NNLO such terms can stem either from fermion box insertions
(light-by-light diagrams) or from fermion triangles insertions where
two fermion legs are cut according to the phase space trigger
function. The latter family vanishes in inclusive observables because
of the Furry's theorem, but they yield a contribution if exclusive phase space cuts
are applied. Light-by-light diagrams were found to have a negligible
numerical impact~\cite{glovervanderbij}, and they were
discarded in the calculation. Thus they are not included in the event
generator {\tt EERAD3} that we use to obtain the fixed-order
distributions. On the other hand, triangle-like diagrams are
numerically sizeable and proportional to $n_F$. Their contribution
does not take part in the running of the coupling so they must not be
absorbed in the scale-fixing procedure.

Looking at Eq.~\eqref{eq:expcoeffs} one can see that by plugging the
LO expression for the $Q^*$ scale into the ${C}(y)$ coefficient, the
$Q^{**}$ dependence in the $n_F^2$ contribution gets cancelled. This
implies that the only way to absorb the $n_F^n$ term at ${\cal
  O}(\alpha_s^{n+1})$ is to modify the LO scale $Q^*$ by radiative
corrections~\cite{Brodsky:1982gc}. It is straightforward to show that
this leads to the choice
\begin{align}
\label{eq:BLMNNLO1-1}
Q^*&= Q \exp \left(\frac{3 B_1(y)}{2 A_0(y)}\right) \notag\\
&\times
\bigg(1+9\frac{\alpha_s(\mu)}{2\pi}\beta_0\left(\frac{B_1^2(y)}{A_0^2(y)}-\frac{C_2(y)}{A_0(y)}\right)+
\ldots\bigg),
\end{align}
where we use the conventions of ref.~\cite{Monni:2011gb} for the QCD
$\beta$ function, {\it i.e.} $\beta_{0} =11/12C_{A}-1/3T_{F}n_{F}$.
It should be noted that we absorb a contribution
  proportional to $\beta_0$, such that the dependence on the number of
  active flavours is fully contained in the coefficients of the QCD
  $\beta$ function.
The scale $\mu$ at which the coupling constant in
Eq.~\eqref{eq:BLMNNLO1-1} is evaluated is determined by higher-order
corrections to the process and it is arbitrary at this order.  It
introduces an intrinsic ambiguity similar to the $Q^{***}$ scale in
the NNLO corrections. The choice~\eqref{eq:BLMNNLO1-1} guarantees the
absence of the leading renormalon $\sim \alpha_s^{n+1}
\beta_0^n n!$ from the perturbative expansion. The NLO scale $Q^{**}$
can be obtained by absorbing the ${\cal O}(\alpha_s^3)$ single vacuum
polarization insertions (proportional to $n_F$) into the coupling. It
is in general very difficult to single out the contributions of
UV-finite diagrams proportional to $n_F$ from the remaining vacuum
polarization terms and the resulting decomposition is not
gauge-invariant. The resulting expression for the $Q^{**}$ scale
reads~\cite{Brodsky:1994eh}
\begin{align}
  \label{eq:BLMNNLO1-2}
  Q^{**} & = Q \exp \bigg \{-\frac{3\left(19 B_1(y)-2C^{\rm VP}_1(y)-66C_2(y)\right)}{4\left(2B_0(y)+33B_1(y)\right)} \bigg \},
\end{align}
where $C^{\rm VP}_1(y)$ contains only the vacuum polarization
contributions to the $C_1(y)$ coefficient. \\
Nevertheless, Eq.~\eqref{eq:BLMNNLO1-2} often leads to very low values
of the scale $Q^{**}$ also in the hard region of the spectrum for
almost all of the observables studied here. The resulting scales are of
the order of $\Lambda_{\rm QCD}$, and this makes the choice in
Eq.~\eqref{eq:BLMNNLO1-2} useless. We thus decide to implement a
minimal prescription where we only set the $Q^*$ scale to its NLO
value (Eq.~\eqref{eq:BLMNNLO1-1}), while keeping $Q^{**}=Q^{***}=\mu=
Q$. We stress that there is no reason why one should set
  $Q^{**}=Q^{*}$, since this would introduce spurious $n_F$-dependent
  terms at ${\cal O}(\alpha_s^3)$.  

\section{Numerical results}
\label{sec:results}
\begin{figure}[htp!]
  \centering
  \includegraphics[width=0.5 \textwidth]{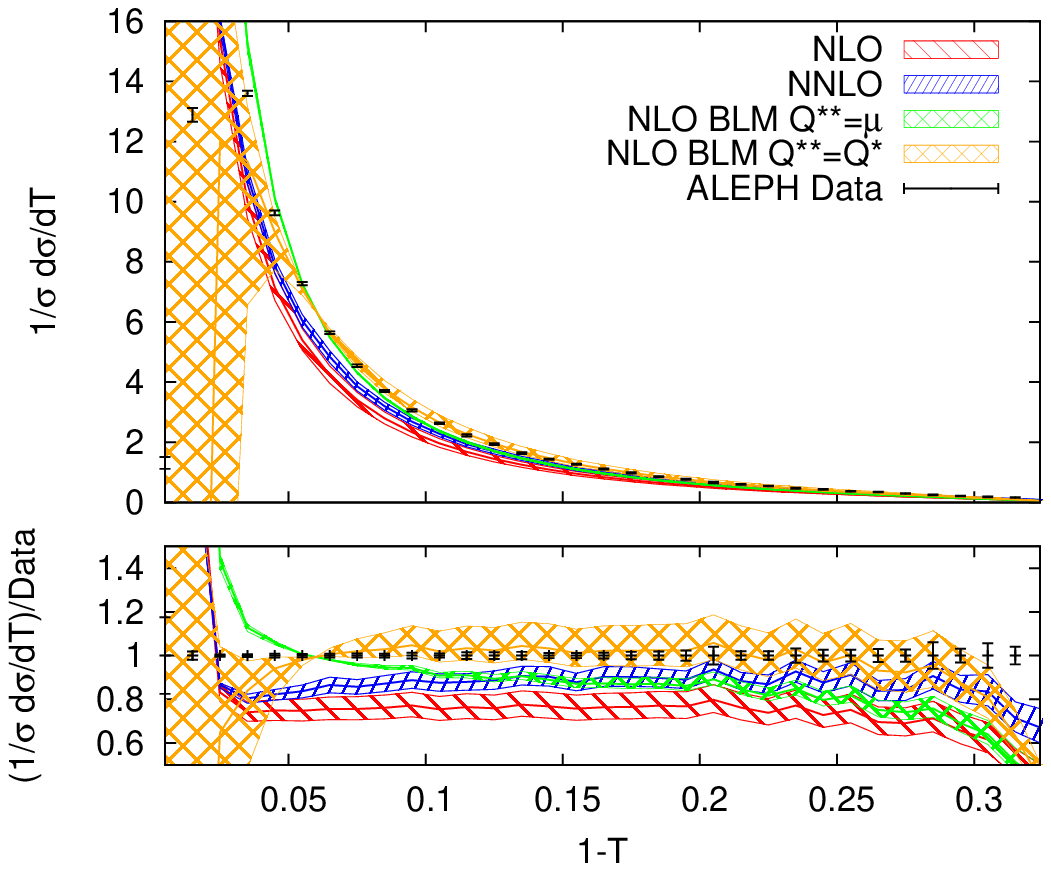}
  \vspace{1em}
  \includegraphics[width=0.5 \textwidth]{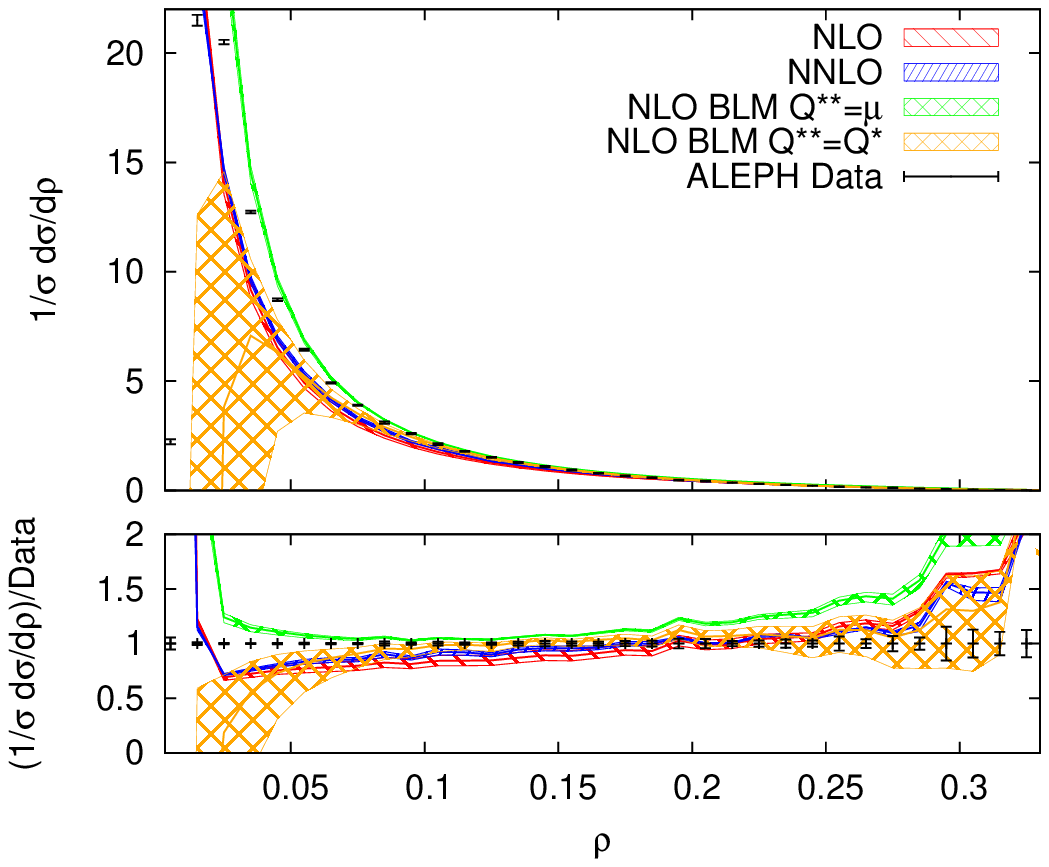}
  \vspace{1em}
  \includegraphics[width=0.5 \textwidth]{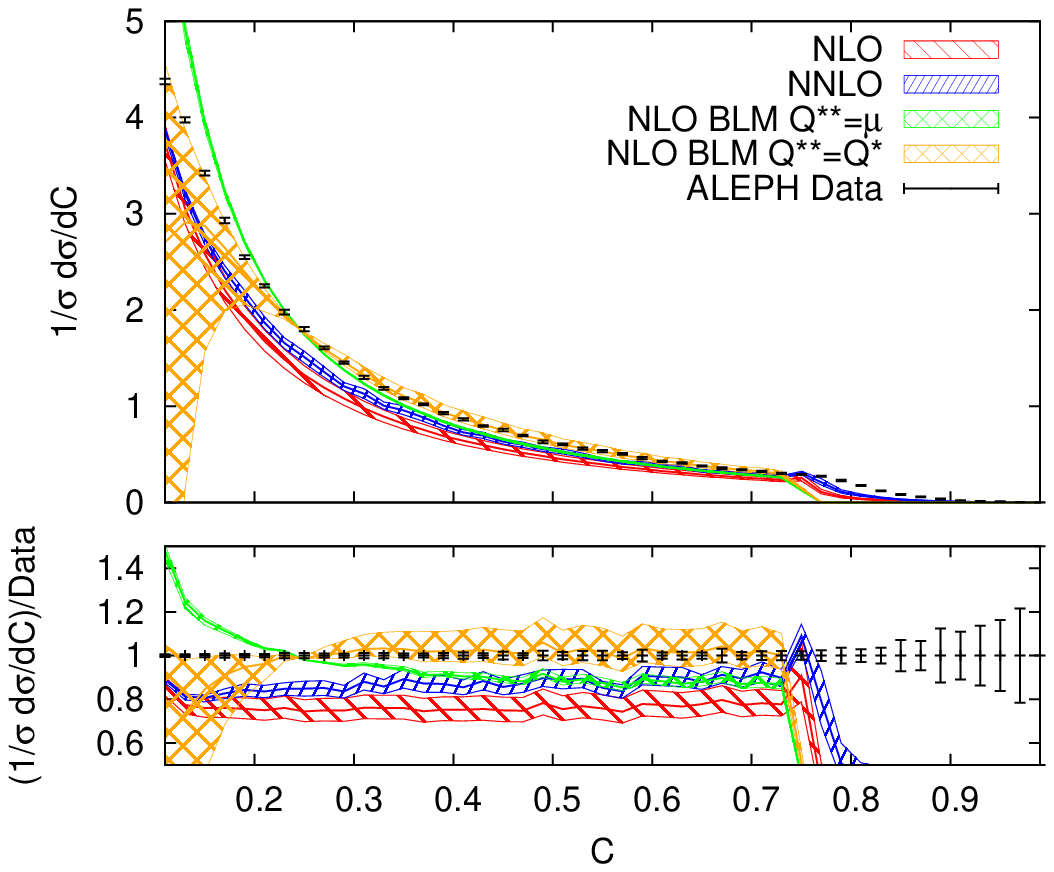} 
  \caption{Distributions for Thrust ($T$), Heavy-jet mass ($\rho = M_H^2/Q^2 $), and
    C-parameter ($C$) at $Q=M_Z$. The red and blue curves are the
    fixed-scale NLO and NNLO predictions, respectively. The remaining
    bands represent the NLO prediction with BLM scale fixing either
    with $Q^{**}=Q^*$ (orange) or $Q^{**}=\mu$
    (green).}\label{fig:blmnlo1}
\end{figure}

\begin{figure}[htp!]
  \centering
  \vspace{0.25em}
  \includegraphics[width=0.5 \textwidth]{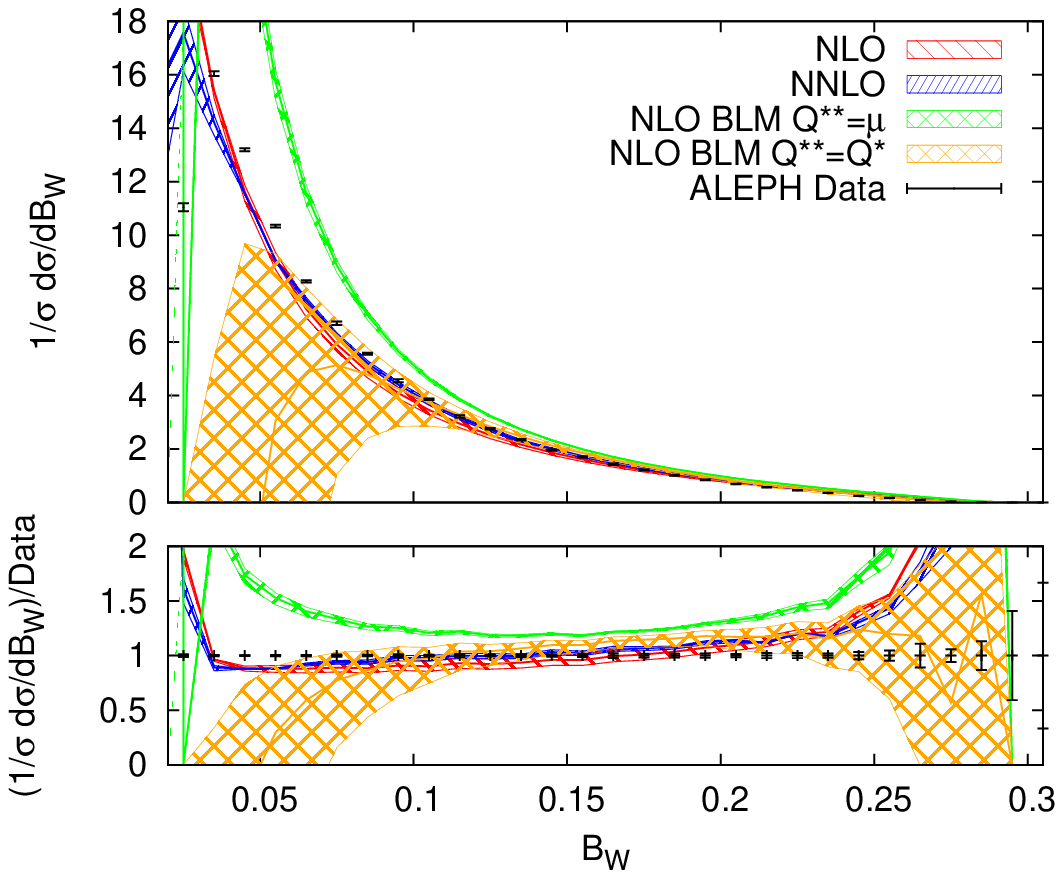}
  \vspace{1em}
  \includegraphics[width=0.5 \textwidth]{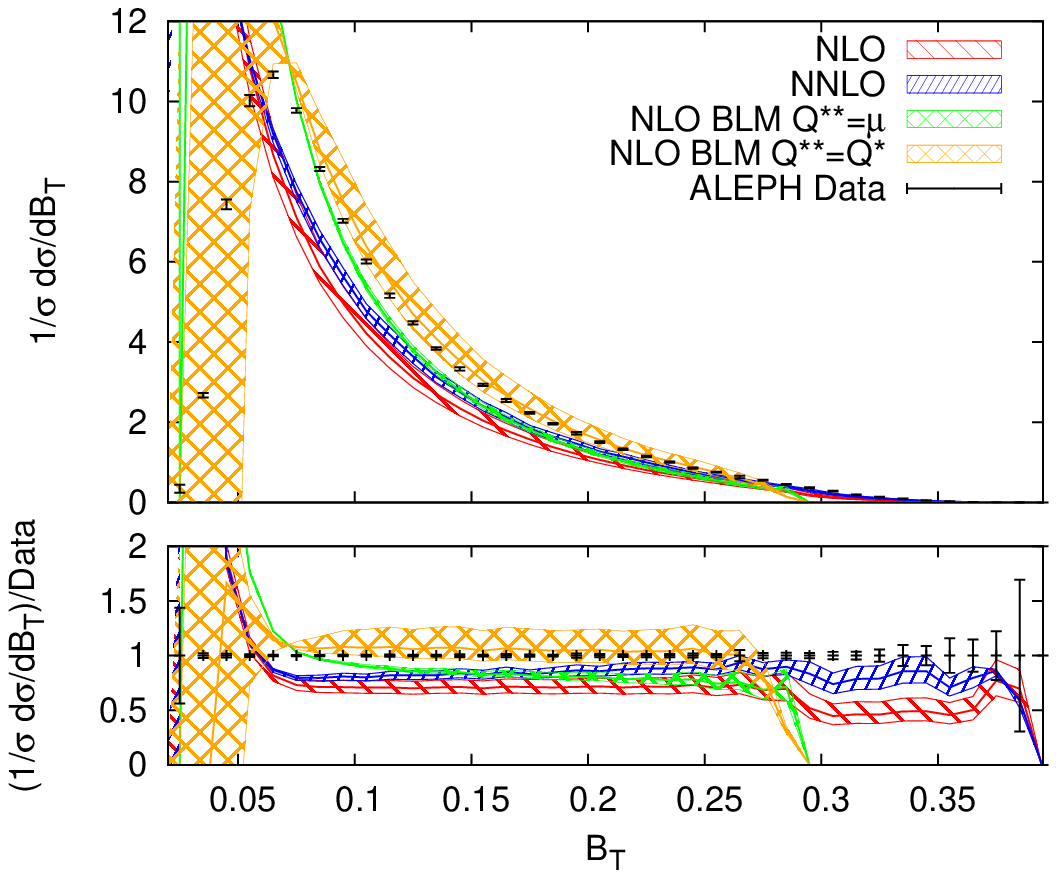} 
  \vspace{1em}
  \includegraphics[width=0.5 \textwidth]{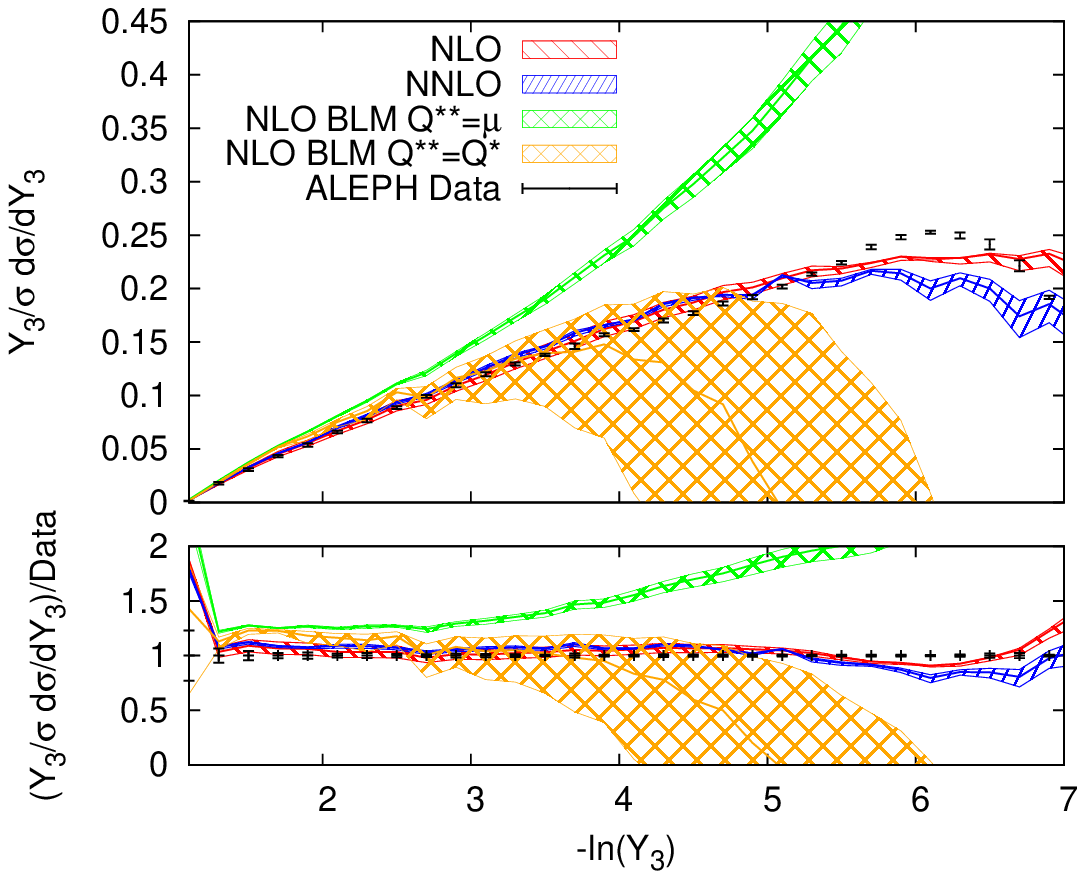}
  \caption{Distributions for total and wide broadening ($B_W$,$B_T$),
    and three-jet resolution parameter in the Durham ($k_t$) algorithm
    ($Y_{3}$) at $Q=M_Z$. The red and blue curves are the fixed-scale
    NLO and NNLO predictions, respectively. The remaining bands
    represent the NLO prediction with BLM scale fixing either with
    $Q^{**}=Q^*$ (orange) or $Q^{**}=\mu$ (green).}\label{fig:blmnlo2}
\end{figure}

In this section we present and discuss the numerical results~\cite{niklausthesis} obtained
with the scale-fixing prescriptions discussed above.  Distributions
are obtained with the generator {\tt EERAD3}, yielding higher order 
coefficients normalized to 
the Born cross section
\begin{equation}
\sigma_0 = {4 \pi \alpha \over 3 s} N e_q^2,
\label{eq:sigma0}
\end{equation}
while the formulae presented in the previous sections are obtained by
normalizing the differential cross sections to the total cross section
$\sigma$ for $e^+e^-\to$\,hadrons. One can account for the change of
normalization by means of the factor
\begin{multline}
{\sigma_0 \over \sigma } = 1 - {3 \over 2} C_F \left( {\alpha_s \over 2 \pi} \right ) + {C_F \over 8} \left (21 C_F+ n_f T_F \left (44 - 32 \zeta(3) \right) \right. \\
+ \left. C_A \left (88 \zeta(3) -123 \right) \right) \left({ \alpha_s \over 2 \pi} \right)^2 + \mathcal{O}(\alpha_s^3).
\end{multline}
The running coupling is evaluated using the package {\tt
  RunDec}~\cite{Chetyrkin:2000yt}.  Figures~\ref{fig:blmnlo1} and
\ref{fig:blmnlo2} show the comparison between the NLO and NNLO
distributions evaluated at a fixed renormalization scale $Q$ (red and
blue curves, respectively) and two different implementations of the
BLM method corresponding to two different choices for the NLO scale
$Q^{**}$. The green curve is obtained by setting $Q^{**}=Q$ while the
orange curve corresponds to $Q^{**}=Q^{*}$. Experimental data from the
ALEPH experiment~\cite{aleph} at $Q=M_Z$ are also included. The error
bands for the standard fixed-order results (red and blue bands) are
obtained by varying the renormalization scale initially set to $Q$ by
a factor of two in either direction. The bands for the BLM curves are
obtained by implementing the latter variation for the $Q^{**}$ scale
around its central value. Such a scale is ambiguous at this order and
its variation gives an estimate of the uncertainty associated with it.
We observe that in both cases the BLM prescription gives rise to a
harder spectrum for all observables. The choice $Q^{**}=Q$ (green
band) leads to a smaller error band when compared to both the NLO and
the NNLO ones. On the other hand, the choice $Q^{**}=Q^*$ (orange
band) leads to much larger errors. The distributions obtained with the
latter choice are in good agreement with experimental data away from
the infrared region. Moreover, when the infrared limit is approached,
the fixed-order prediction becomes unreliable and the uncertainty band
gets wider. The BLM method cannot be defined in the multijet region
beyond the leading-oder kinematical endpoint at which the $A_0(y)$
coefficient vanishes, leaving the $Q^*$ scale undefined.

\begin{figure}[htp!]
  \centering
  \includegraphics[width=0.5 \textwidth]{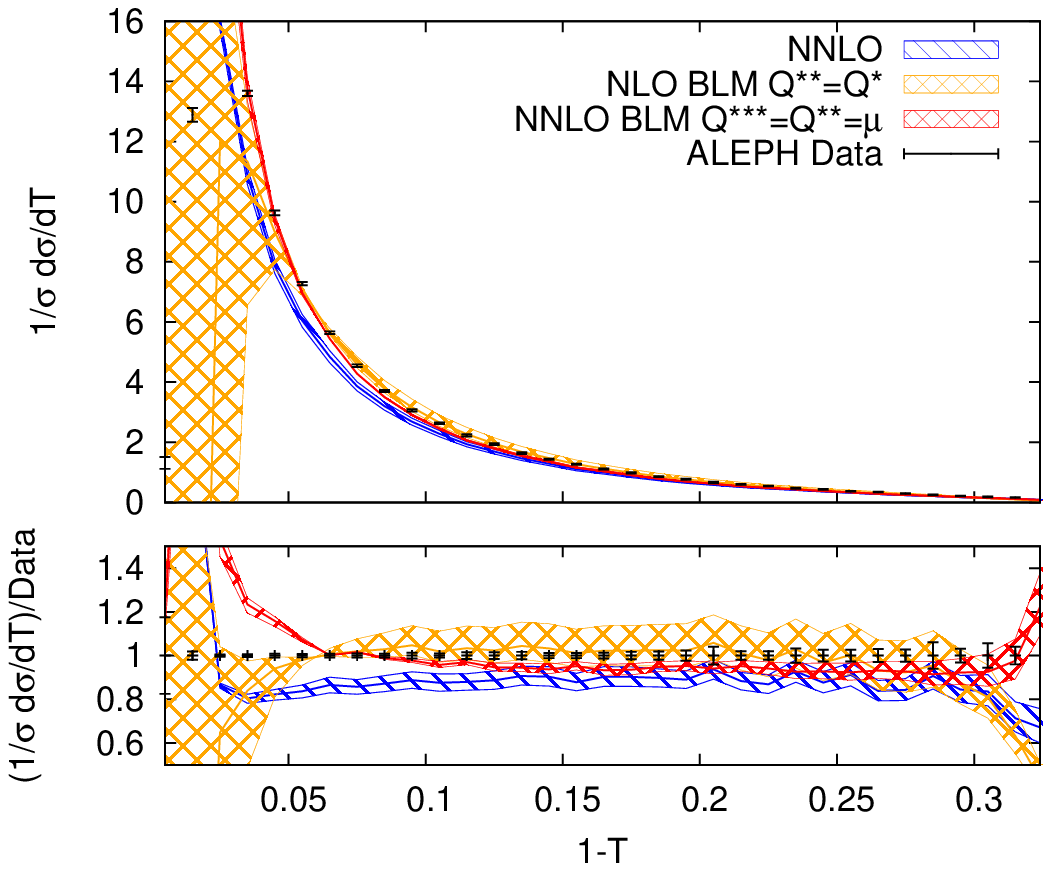}
  \vspace{1em}
  \includegraphics[width=0.5 \textwidth]{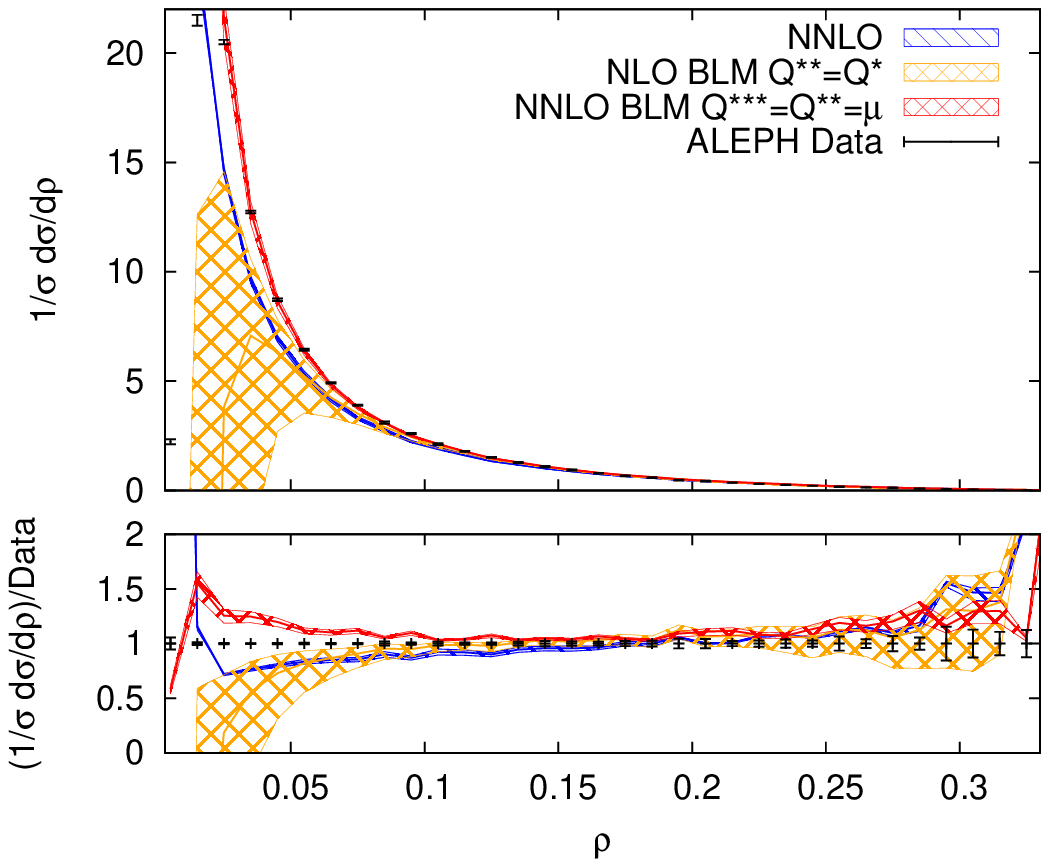}
  \vspace{1em}
  \includegraphics[width=0.5 \textwidth]{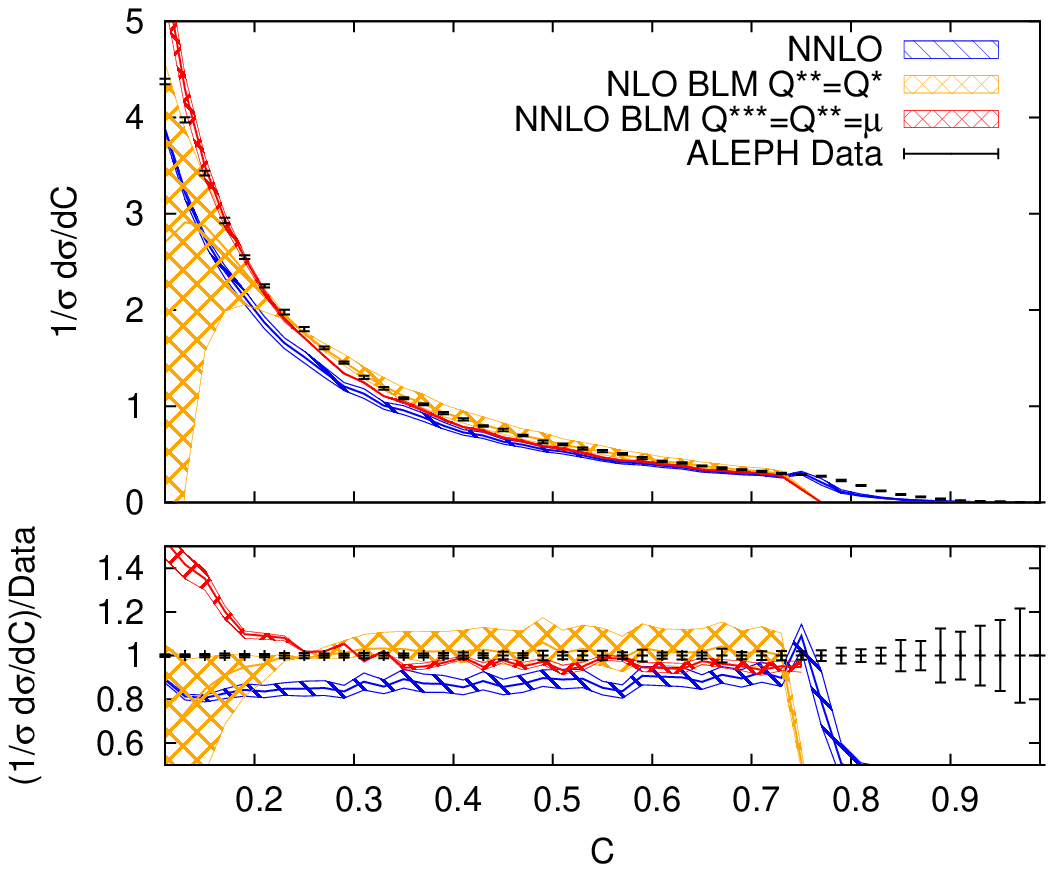} 
  \caption{Distributions for thrust ($T$), heavy-jet mass ($\rho = M_H^2/Q^2$), and
    $C$-parameter ($C$) at $Q=M_Z$. The red and blue curves represent
    the NNLO predictions, either with fixed renormalization scale
    (blue) or with the minimal extension of the BLM method described
    in the text (red). The orange band represents the NLO prediction
    with BLM scale fixing with $Q^{**}=Q^*$.}\label{fig:blmnnlo1}
\end{figure}

\begin{figure}[htp!]
  \centering
  \vspace{0.25em}
  \includegraphics[width=0.5 \textwidth]{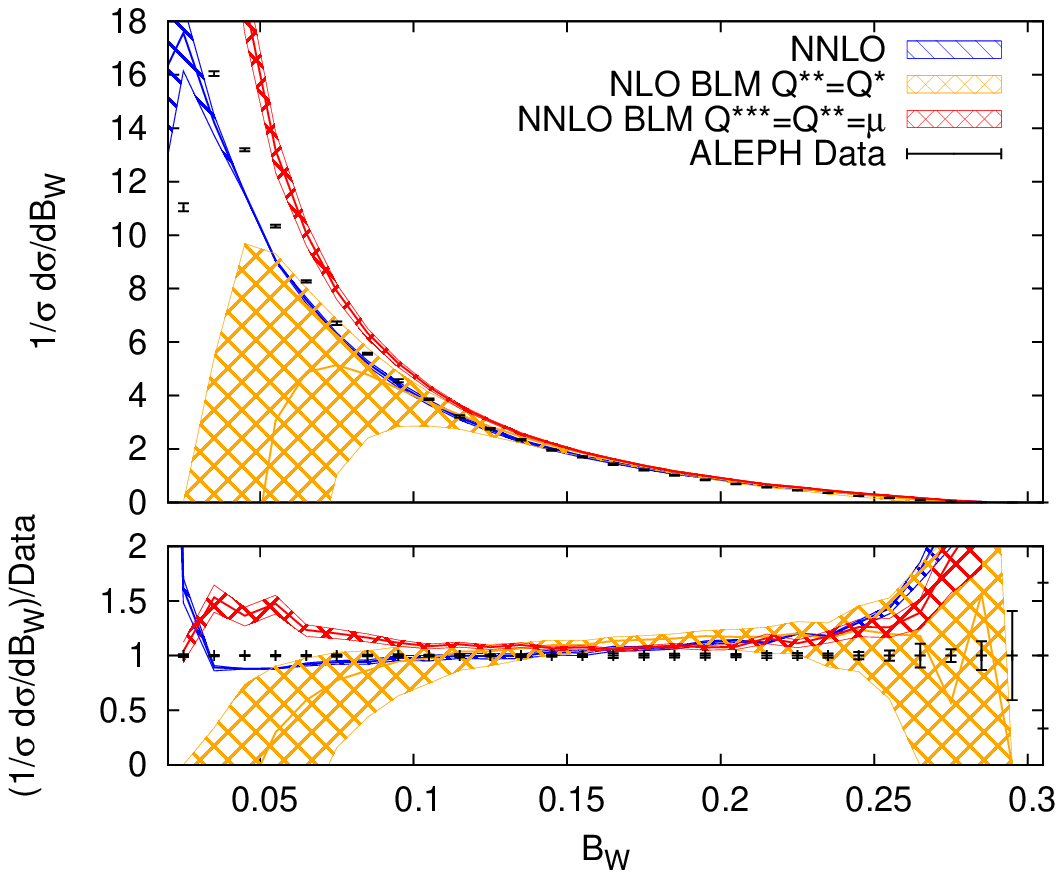}
  \vspace{1em}
  \includegraphics[width=0.5 \textwidth]{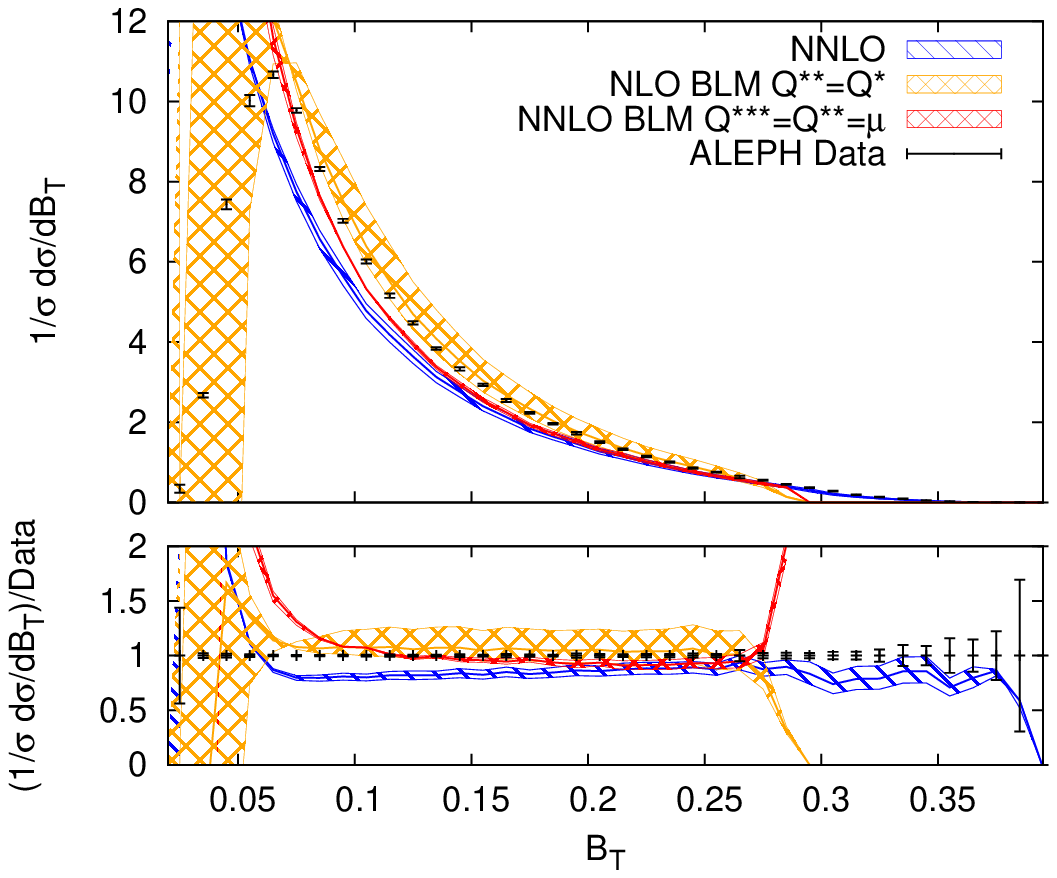} 
  \vspace{1em}
  \includegraphics[width=0.5 \textwidth]{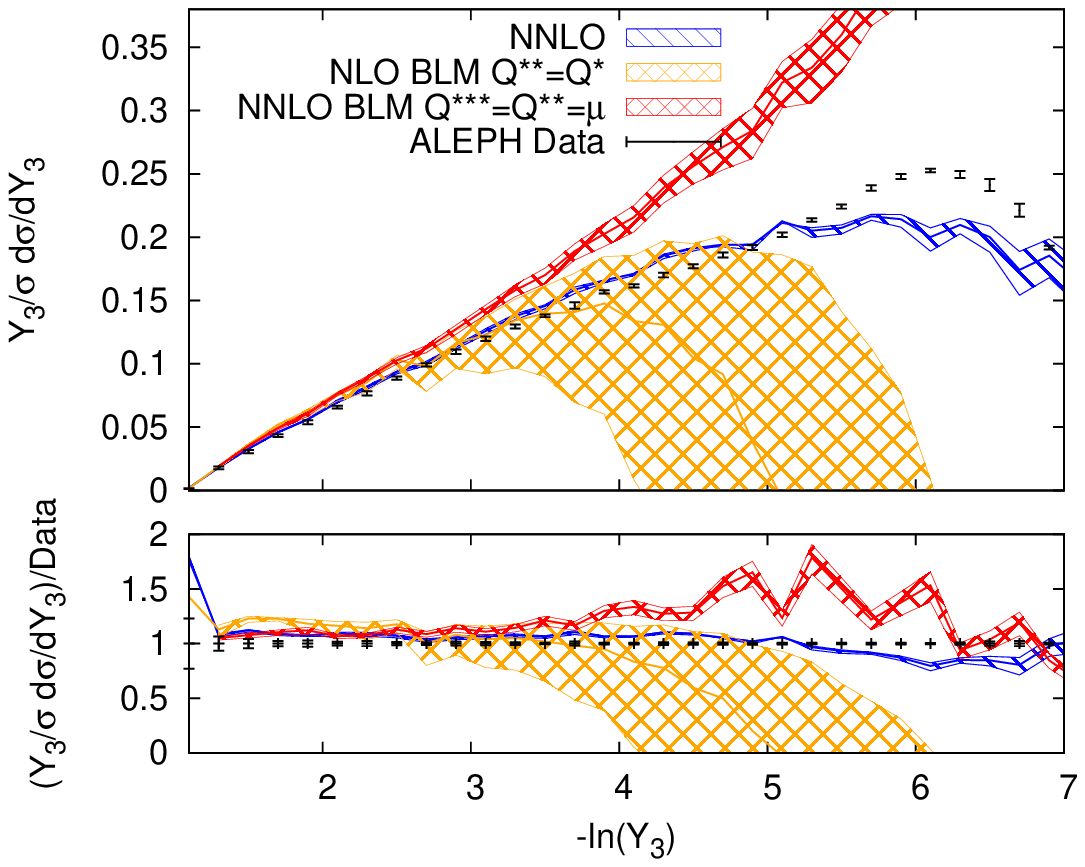}
  \caption{Distributions for wide and total  broadening ($B_W$, $B_T$),
    and three-jet resolution parameter in the Durham algorithm
    ($Y_{3}$) at $Q=M_Z$. The red and blue curves represent the NNLO
    predictions, either with fixed renormalization scale (blue) or
    with the minimal extension of the BLM method described in the text
    (red). The orange band represents the NLO prediction with BLM
    scale fixing with $Q^{**}=Q^*$.}\label{fig:blmnnlo2}
\end{figure}

Figures~\ref{fig:blmnnlo1} and \ref{fig:blmnnlo2} show the comparison
between the NNLO distributions obtained with different scale-fixing
prescriptions.  The blue band corresponds to the standard choice
$Q^*=Q^{**}=Q^{***}=Q$ for the central scale. Its uncertainty is
obtained by varying simultaneously all scales by a factor of two
around $Q$.  The red band is obtained with $Q^*$ set to its NLO value
(Eq.~\eqref{eq:BLMNNLO1-1}) and $Q^{**}=Q^{***}=\mu=Q$, and the
uncertainty is obtained by varying the latter three scales by a factor
of two in either direction. The orange band represents the NLO result
with BLM scale $Q^{**}= Q^*$ discussed above. We observe that the red
curve is pushed towards data and the resulting spectrum is harder. The
corresponding uncertainties are quite small and comparable to the
fixed-scale NNLO ones. It is however very difficult to estimate the
perturbative uncertainty using the BLM prescription due to the
different renormalization scales which enter at different orders.  We
observe that our prescription gives a good description of experimental
data for Thrust, $C$-parameter, heavy-jet mass, and total jet
broadening whilst it fails in the case of the wide jet broadening and
the three-jet resolution parameter.

\begin{figure}[htp!]
  \centering
  \includegraphics[width=0.5 \textwidth]{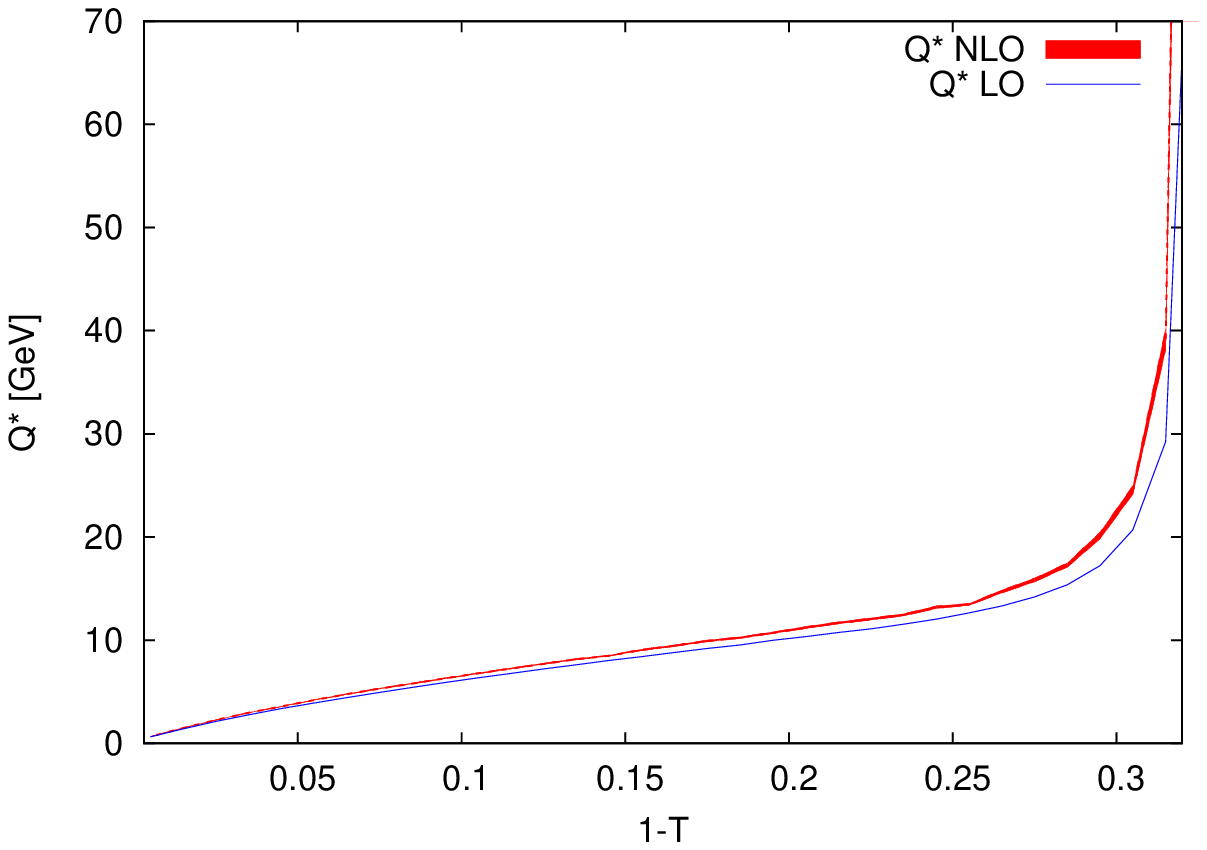}
  \vspace{1em}
  \includegraphics[width=0.5 \textwidth]{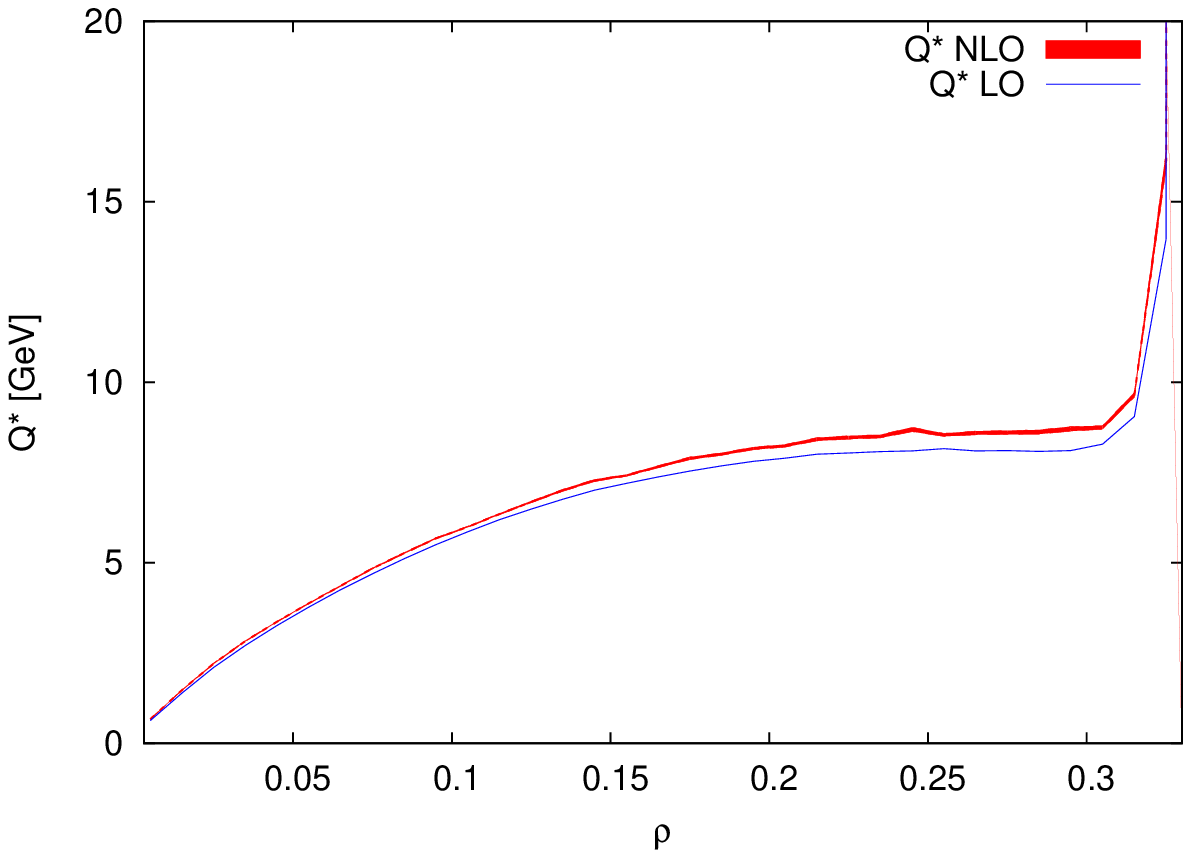}
  \vspace{1em}
  \includegraphics[width=0.5 \textwidth]{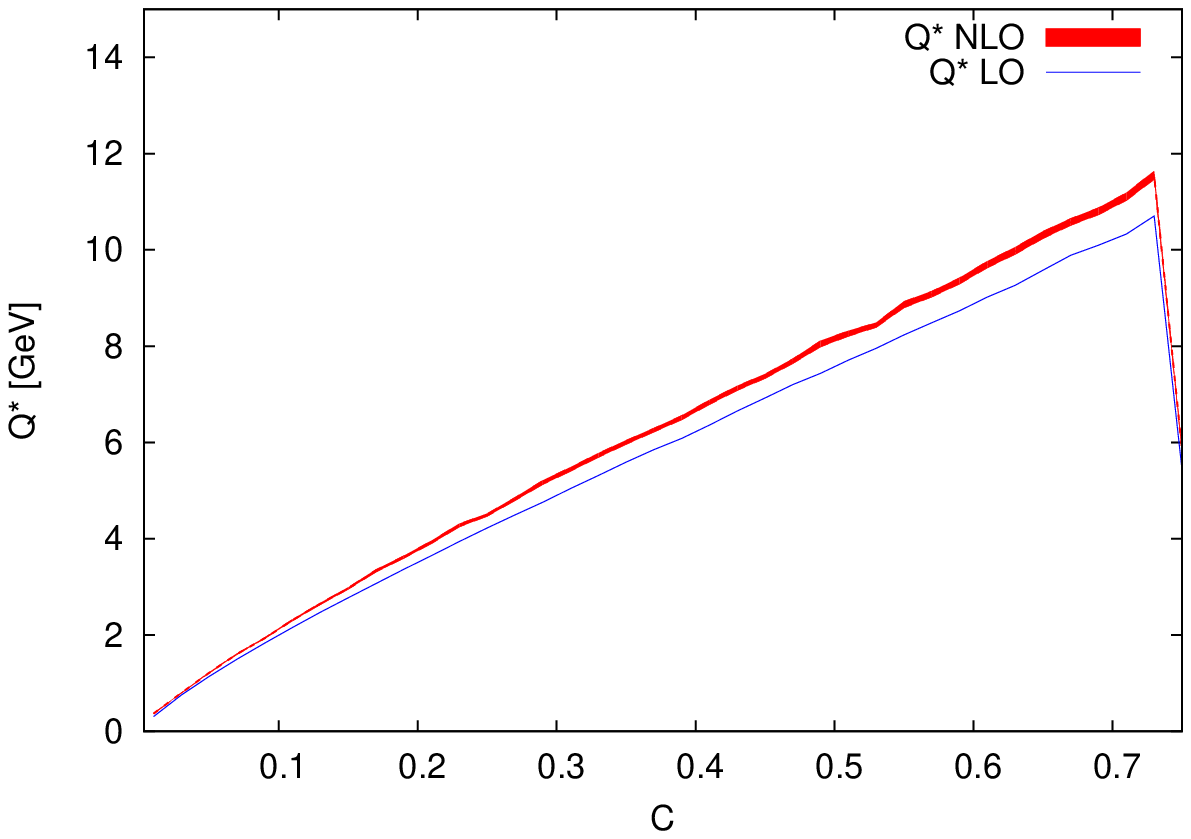} 
  \caption{The LO and NLO BLM scales for thrust ($T$), heavy-jet mass
    ($\rho=M_H^2/Q^2$), and $C$-parameter ($C$).}\label{fig:Qstar1}
\end{figure}

\begin{figure}[htp!]
  \centering
  \vspace{0.25em}
  \includegraphics[width=0.5 \textwidth]{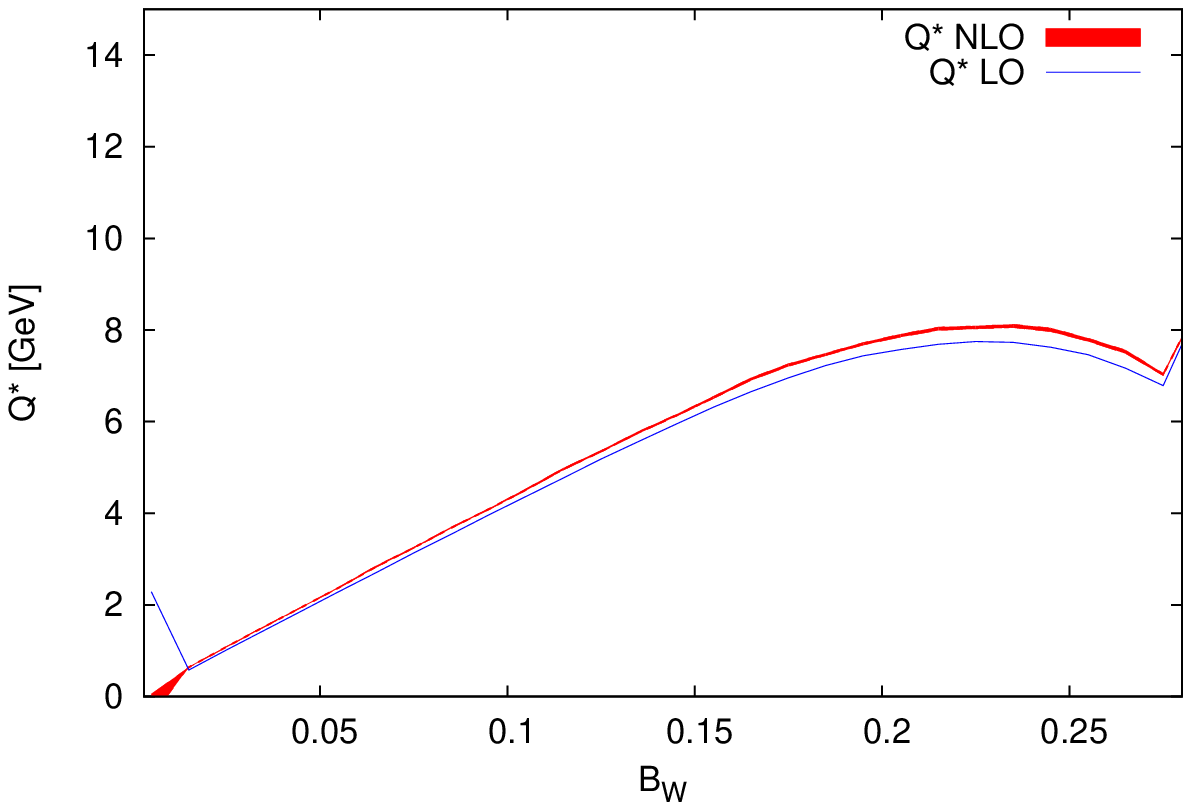}
  \vspace{1em}
  \includegraphics[width=0.5 \textwidth]{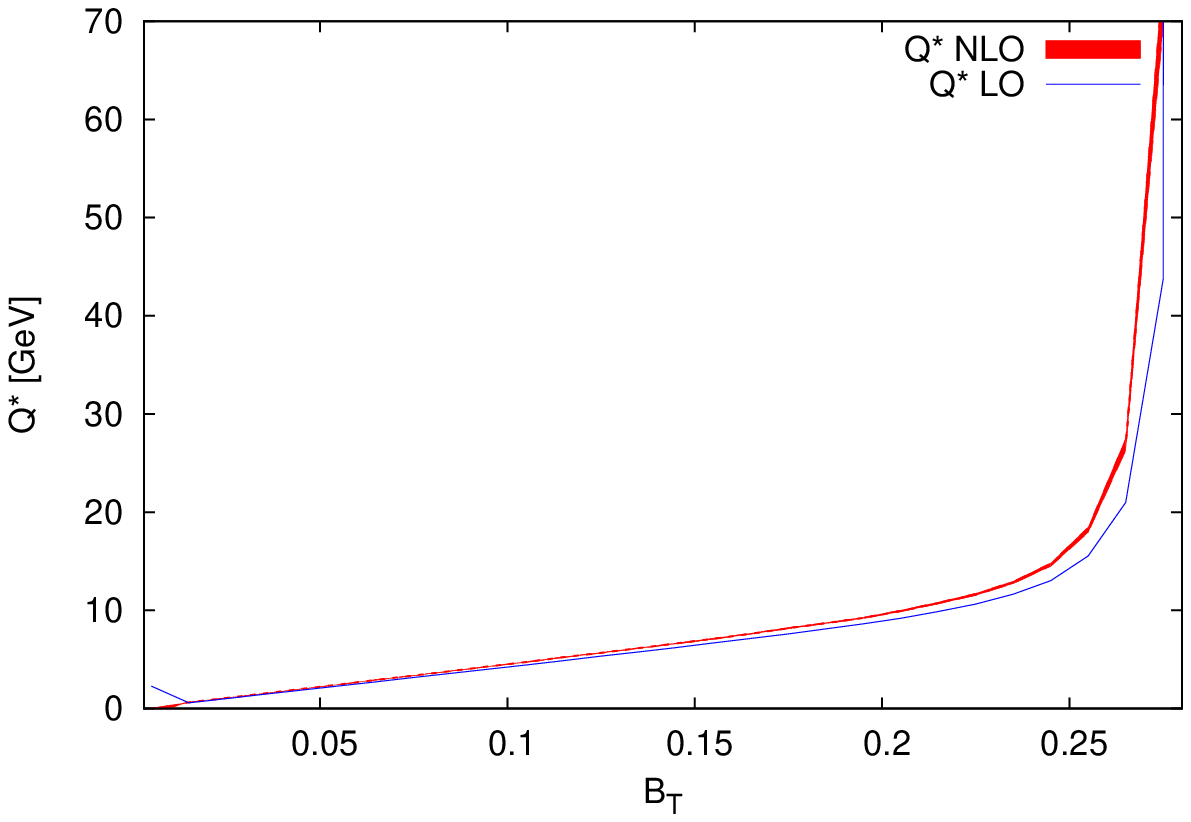} 
  \vspace{1em}
  \includegraphics[width=0.5 \textwidth]{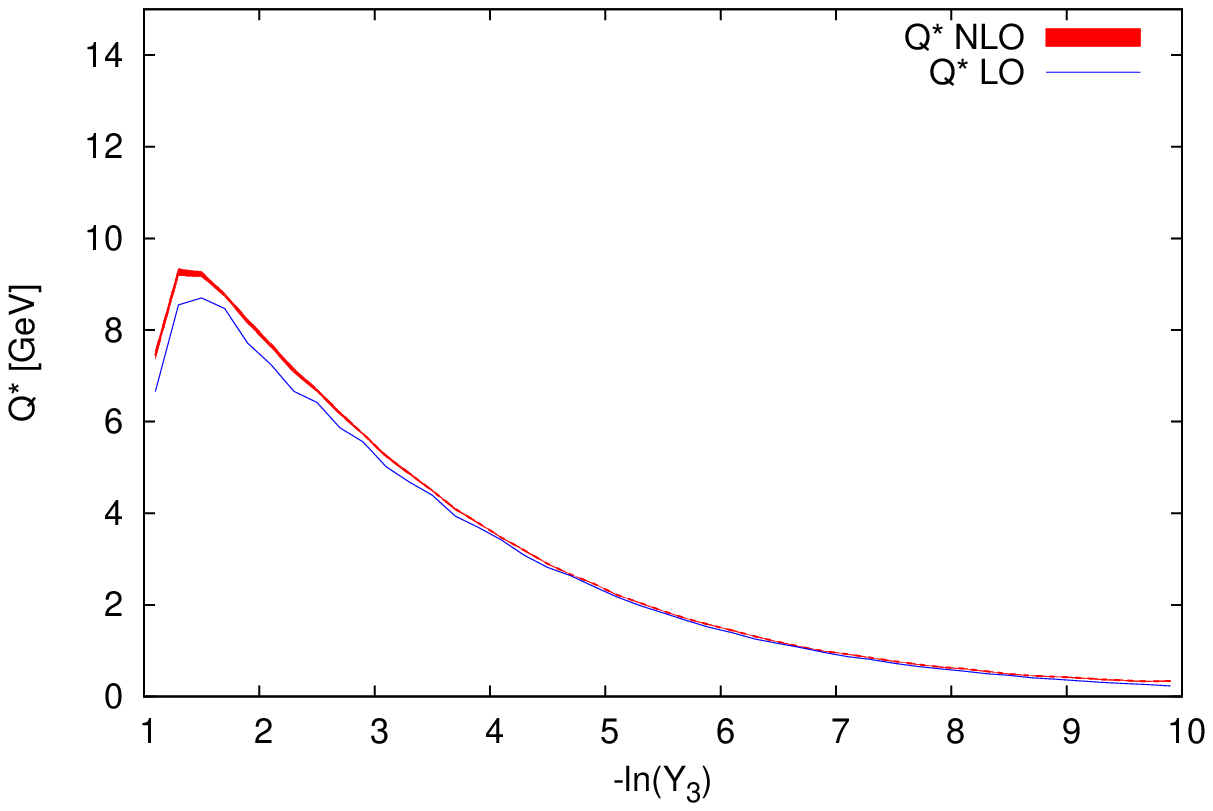}
  \caption{The LO and NLO BLM scales for wide and total broadening
    ($B_W$, $B_T$), and three-jet resolution parameter in the Durham
    algorithm ($Y_{3}$).}\label{fig:Qstar2}
\end{figure}

It is interesting to look at the way the BLM scales behave along the
event-shape spectrum. We plot both the LO and NLO $Q^*$ scale in
Figures~\ref{fig:Qstar1} and \ref{fig:Qstar2}. The blue curve represents
the leading order value, which is independent of any renormalization
scale, while its NLO values are spanned by the red band obtained
by varying the scale $\mu$ in Eq.~\eqref{eq:BLMNNLO1-1} by a factor of
two around $\mu=Q$. We observe that the BLM method leads
to very low renormalization scales, much smaller than the
centre-of-mass energy of the process. The smaller renormalization
scales, thus the larger coupling, lead to harder distributions as
observed above. Moreover, we see that the NLO corrections to the BLM
scale $Q^*$ are quite moderate for all observables, and the radiative
corrections are always positive. In ref.~\cite{Brodsky:1994eh} the BLM
method is applied to the perturbative expansion of inclusive physical
quantities. The authors exponentiate the NLO corrections to the $Q^*$
scale in order to obtain a positive definite quantity. Its expression
reads
\begin{eqnarray}
\label{eq:BLMNNLO1-3}
Q^*&= &Q\exp\bigg(\frac{3 B_1(y)}{2 A_0(y)}
\nonumber \\
& & \hspace{5mm}+9\frac{\alpha_s(\mu)}{2\pi}\beta_0\left(\frac{B_1^2(y)}{A_0^2(y)}-\frac{C_2(y)}{A_0(y)}\right)+\ldots  \bigg).
\end{eqnarray}

Nevertheless, in our case this is not a good approximation of the
correct $Q^*$ scale due to the moderately large NLO corrections. In
practice, the two results Eq.\eqref{eq:BLMNNLO1-1} and
Eq.\eqref{eq:BLMNNLO1-3} lead to very different numerical values for
$Q^*$.

\section{Conclusions}
\label{sec:Conclusions}
In this paper we studied the impact of the BLM scale-setting method on
event-shape distributions in electron-positron collisions. We found
good agreement between the NLO prediction with $Q^{**}=Q^*$ and
ALEPH experimental data at the $Z$-boson peak. The theoretical
uncertainties associated with the latter predictions are larger than
the ones associated with the fixed-scale distributions. 

We also analyzed the extension of the prescription beyond NLO and
found that the scale $Q^{**}$ cannot be defined in a gauge-invariant
manner for the differential cross sections studied here. This is due
to the presence of UV-finite, and $n_F$-dependent terms already at
${\cal O}(\alpha_s^3)$. Hence, the prescription suggested
in~\cite{Brodsky:1994eh} is in general not well-defined for
non-inclusive quantities for which such terms are present. Moreover,
the resulting NLO scale $Q^{**}$ assumes very low values and often
probes the non-perturbative regime of the strong coupling constant. We
therefore implement a minimal prescription in which we set $Q^*$ to
its NLO value~\eqref{eq:BLMNNLO1-1} while setting
$Q^{**}=Q^{***}=\mu=Q$. This prescription ensures the absence of the
leading renormalon $\sim \alpha_s^{n+1} \beta_0^n n!$ ambiguity (up to
higher-order corrections) at a given order in the perturbative
expansion. The agreement with experimental data away from the Sudakov
region is remarkable for all observables but the wide-jet broadening
and the three-jet resolution parameter. We observe a scale uncertainty
of roughly the same size of the fixed-scale NNLO one, and the
resulting distributions are harder.  The renormalization scales
obtained with the BLM method are quite small (of the order of
$10$-$20$\,GeV in the hard region of the spectrum) and radiative
corrections to their value are moderate.

\section{Acknowledgements}
This work was supported by the Swiss National Science Foundation
(SNF) under grant 200020-138206 and the European Commission through
the LHCPhenoNet network under contract PITN-GA-2010-264564.


\clearpage




\end{document}